\documentclass[letter,bibyear]{aa} 

\usepackage{epsfig}
\usepackage{amsmath}
\usepackage{subfigure}
\usepackage{natbib}
\usepackage{multicol}
\usepackage{multirow}
\usepackage{color}
\usepackage{float}
\usepackage{longtable}
\usepackage{graphicx}
\usepackage{epstopdf}
\usepackage{booktabs}
\usepackage{txfonts}

\newcommand{\kms}{km\,s$^{-1}$}
\newcommand{\acsesc}{ACS Earth Space Chem.}

\newcommand{\prev}{Phys. Rev.}

\begin{document}

\title{Laboratory and astronomical discovery of the cyanovinyl radical H$_2$CCCN \thanks{Based on observations with the 40-m radio telescope (projects 19A003, 20A014, 20D023, 21A011, and 21D005) of the National Geographic Institute of Spain (IGN) at Yebes Observatory. Yebes Observatory thanks the ERC for funding support under grant ERC-2013-Syg-610256-NANOCOSMOS.
}}

\author{
C.~Cabezas\inst{1},
J.~Tang\inst{2},
M.~Ag\'undez\inst{1},
K.~Seiki\inst{3},
Y.~Sumiyoshi\inst{4},
Y.~Ohshima\inst{5},
B.~Tercero\inst{6,7},
N.~Marcelino\inst{6,7},
R.~Fuentetaja\inst{1},
P.~de~Vicente\inst{7},
Y.~Endo\inst{8},
J.~Cernicharo\inst{1}
}

\institute{Departamento de Astrof\'isica Molecular, Instituto de F\'isica Fundamental (IFF-CSIC),
C/ Serrano 121, 28006 Madrid, Spain\\
\email carlos.cabezas@csic.es, jtang@okayama-u.ac.jp, jose.cernicharo@csic.es
\and Institute of Global Human Resource Development and Graduate School of Natural Science and Technology, Okayama University, 3-1-1 Tsushima-naka, Kita-ku, Okayama 700-8531, Japan
\and Department of Basic Science, The University of Tokyo, 3-8-1 Komaba, Meguro-ku,Tokyo 153-8902, Japan
\and Division of Pure and Applied Science, Graduate School of Science and Technology, Gunma University, 4-2 Aramaki, Maebashi, Gunma 371-8510, Japan
\and Department of Chemistry, School of Science, Tokyo Institute of Technology, Ookayama 2-12-1, Meguro, Tokyo 152-8550, Japan
\and Observatorio Astron\'omico Nacional (IGN), C/ Alfonso XII 3, 28014 Madrid, Spain
\and Observatorio de Yebes (IGN), Cerro de la Palera s/n, 19141 Yebes, Guadalajara, Spain
\and Department of Applied Chemistry, Science Building II, National Yang Ming Chiao Tung University, 1001 Ta-Hsueh Rd., Hsinchu 300098, Taiwan
}
\date{Received; accepted}

\abstract{ We report the first laboratory and interstellar detection of the $\alpha$-cyano vinyl radical (H$_2$CCCN). This species was produced in the laboratory by an electric discharge of a gas mixture of vinyl cyanide, CH$_2$CHCN, and Ne, and its rotational spectrum was characterized using a Balle-Flygare narrowband-type Fourier-transform microwave spectrometer operating in the frequency region of 8-40 GHz. The observed spectrum shows a complex structure due to tunneling splittings between two torsional sublevels of the ground vibronic state, $0^+$ and $0^-$, derived from a large-amplitude inversion motion. In addition, the presence of two equivalent hydrogen nuclei makes necessary to discern between ortho- and para-H$_2$CCCN. A least squares analysis reproduces the observed transition frequencies with a standard deviation of ca. 3 kHz. Using the laboratory predictions, this radical is detected in the cold dark cloud TMC-1 using the Yebes 40m telescope and the QUIJOTE$^1$ line survey. The 4$_{0,4}$-3$_{0,3}$ and 5$_{0,5}$-4$_{0,4}$ rotational transitions, composed of several hyperfine components, were observed in the 31.0-50.4 GHz range. Adopting a rotational temperature of 6\,K we derive a column density of (1.4$\pm$0.2)$\times$10$^{11}$ cm$^{-2}$ and (1.1$\pm$0.2)$\times$10$^{11}$ cm$^{-2}$ for ortho-H$_2$CCCN and para-H$_2$CCCN, respectively. The reactions C + CH$_3$CN, and perhaps also N + CH$_2$CCH, emerge as the most likely routes to H$_2$CCCN in TMC-1.}


\titlerunning{Discovery of H$_2$CCCN in TMC-1}
\authorrunning{Cabezas et al.}

\maketitle

\section{Introduction}

Vinyl cyanide, CH$_2$CHCN, is a well known interstellar molecule that was detected for the first time in the interstellar medium (ISM) in 1973 toward the Sagittarius B2 (Sgr B2) molecular cloud \citep{Gardner1975}. Since then, vinyl cyanide has been detected toward different sources, such as Orion \citep{Schilke1997}, the dark cloud TMC-1 \citep{Matthews1983}, the circumstellar envelope of the late-type star IRC+10216 \citep{Agundez2008}, and the Titan atmosphere \citep{Capone1981}. Vinyl cyanide is one of the molecules, whose high abundance and significant dipole moment allow radioastronomical detection even of its rare isotopologue species and vibrationally
excited states  \citep{Lopez2014}. Thus, the cyanovinyl radical (CVR) is a promising candidate for its interstellar detection given its similarity to the aforementioned vinyl cyanide. This hypothesis is strengthened by the recent detection of the H$_2$C$_4$N radical \citep{Cabezas2021} with the QUIJOTE\footnote{\textbf{Q}-band \textbf{U}ltrasensitive \textbf{I}nspection \textbf{J}ourney
to the \textbf{O}bscure \textbf{T}MC-1 \textbf{E}nvironment} line survey \citep{Cernicharo2021,Cernicharo2023}.

The CVR has two structural isomers depending on whether the $\alpha$- or $\beta$-hydrogen, -CH or CH$_2$, respectively, of vinyl cyanide is removed. The molecular formula H$_2$CCCN corresponds to the $\alpha$-CVR while HCCHCN formula is that for $\beta$-CVR, which further allows the cis and trans isomers. Quantum chemical calculation indicate that the $\alpha$-CVR is more stable in energy than the $\beta$-CVR by 30–45 kJ mol$^{-1}$ \citep{Balucani2000,Huang2000,Johansen2019}. The rotational spectra of both isomers of
the $\beta$-CVR have been investigated by \citet{Johansen2019} and \citet{Nakajima2022}. In their study, \citet{Johansen2019} observed several rotational transitions for both cis and trans isomers of $\beta$-CVR in the 5-75 GHz frequency range but no conclusive assignments of the fine and hyperfine components was reported. On the other hand, \citet{Nakajima2022} reported precise molecular constants for the cis isomer of $\beta$-CVR thanks to the deep analysis of all the fine and hyperfine components observed in the 10-55
GHz region. However, no spectroscopic data for the $\alpha$-CVR have been reported before in the literature, except our preliminary results \citep{Tang2000}.

In this Letter we report the first rotational investigation study of the $\alpha$-CVR, hereafter H$_2$CCCN, and the discovery of this radical in space towards TMC-1. The laboratory characterization has been done using Fourier transform microwave (FTMW) spectroscopy in combination to electric discharges techniques. The identification of this radical in space has been done using the on-going QUIJOTE line survey \citep{Cernicharo2021,Cernicharo2023}.

\section{Laboratory FTMW spectroscopy of H$_2$CCCN} \label{labo_observations}

The rotational spectrum of H$_2$CCCN was observed using a Balle-Flygare narrowband type FTMW spectrometer operating in the frequency region of 4-40 GHz \citep{Endo1994,Cabezas2016}. The short-lived species H$_2$CCCN was produced in a supersonic expansion by a pulsed electric discharge of a gas mixture of CH$_2$CHCN (0.2\%) diluted in Ne. This gas mixture was flowed through a pulsed-solenoid valve that is accommodated in the backside of one of the cavity mirrors and aligned parallel to the optical axis of the resonator. A pulse voltage of 900 V with a duration of 450 $\mu$s was applied between stainless steel electrodes attached to the exit of the pulsed discharge nozzle (PDN), resulting in an electric discharge synchronized with the gas expansion.
The resulting products generated in the discharge were supersonically expanded, rapidly cooled to a rotational temperature of $\sim$2.5\,K between the two mirrors of the Fabry-P\'erot resonator, and then probed by FTMW spectroscopy. For measurements of the paramagnetic lines, the Earth’s magnetic field was cancelled by using three sets of Helmholtz coils placed perpendicularly to one another. Since the PDN is arranged parallel to the cavity of the spectrometer, it is possible to suppress the Doppler broadening of the spectral lines, allowing to resolve small hyperfine splittings. The spectral resolution is 5\,kHz and the frequency measurements have an estimated accuracy better than 3\,kHz.

\begin{figure}
\centering
\includegraphics[angle=0,width=0.5\columnwidth]{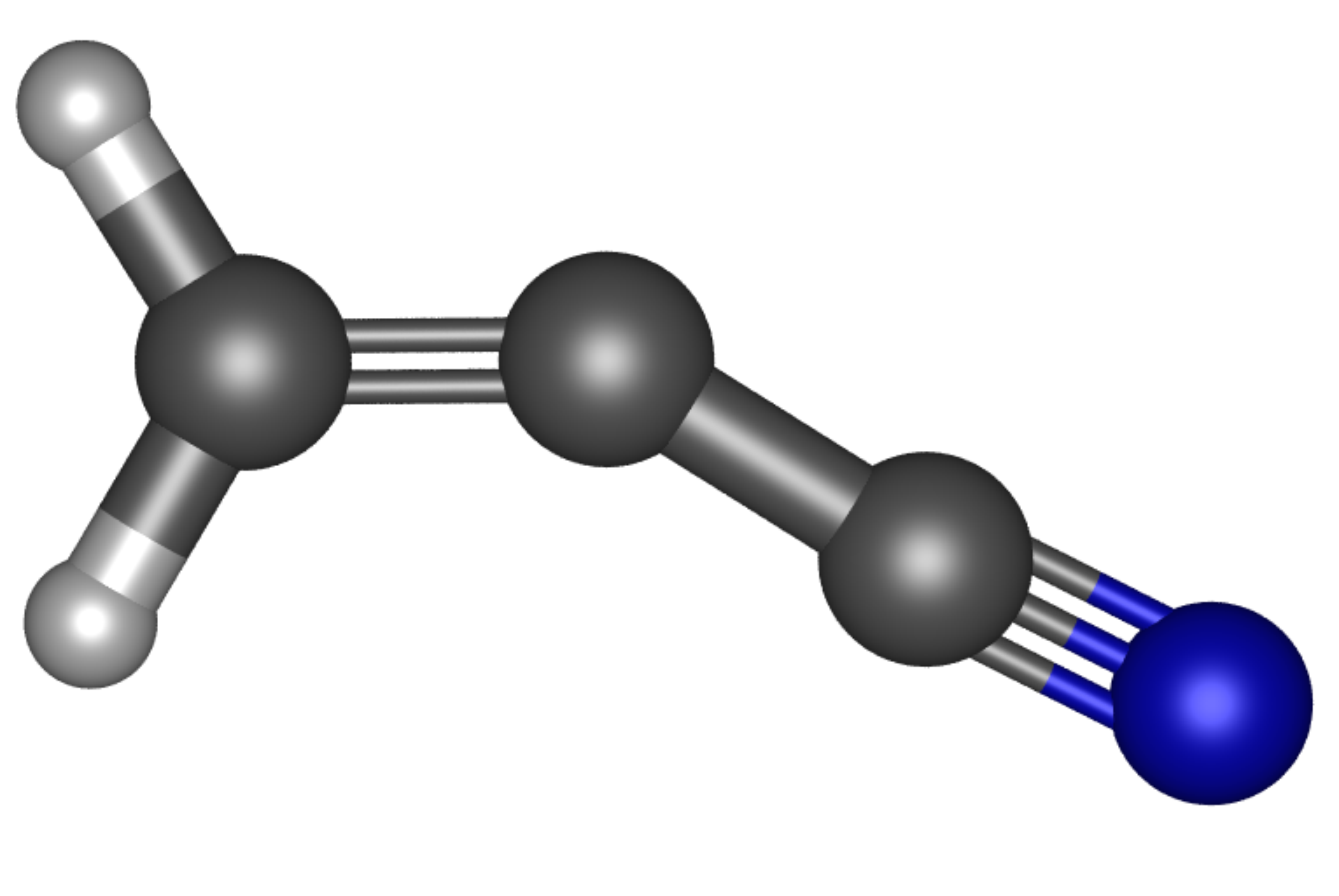}
\caption{Molecular structure of the H$_2$CCCN radical.}
\label{structure}
\end{figure}

\section{Astronomical observations} \label{astro_observations}

New receivers, built within the Nanocosmos\footnote{ERC grant ERC-2013-Syg-610256-NANOCOSMOS.\\
https://nanocosmos.iff.csic.es/} project and installed at the Yebes 40 m radiotelescope, were used for the observations of TMC-1 ($\alpha_{J2000}=4^{\rm h} 41^{\rm  m} 41.9^{\rm s}$ and $\delta_{J2000}= +25^\circ 41' 27.0''$). A detailed description of the telescope, receivers, and backends is given by \citet{Tercero2021}. Briefly, the receiver consists of two cold high electron mobility transistor amplifiers covering the 31.0-50.3 GHz band with horizontal and vertical polarizations. The backends are $2\times8\times2.5$ GHz fast Fourier transform spectrometers with a spectral resolution of 38.15 kHz providing the whole coverage of the Q-band in both polarisations.
The observations, carried out during different observing runs, are performed using the frequency-switching mode with a frequency throw of 10 MHz in the very first observing runs, during November 2019 and February 2020, 8 MHz during the observations of January-November 2021, and alternating these frequency throws in the last observing runs between October 2021 and February 2023. The total on-source telescope time is 850 hours in each polarization (385 and 465 hours for the 8 MHz and 10 MHz frequency throws, respectively). The sensitivity of the QUIJOTE line survey varies between 0.09 and 0.25\,mK in the 31-50.3\,GHz domain. The intensity scale used in this work, antenna temperature ($T_A^*$), was calibrated using two absorbers at different temperatures and the atmospheric transmission model ATM \citep{Cernicharo1985,Pardo2001}. Calibration uncertainties have been adopted to be 10~\%. The beam efficiency of the Yebes 40 m telescope in the Q-band is given as a function of frequency by $B_{\rm eff}$= 0.797 exp[$-$($\nu$(GHz)/71.1)$^2$]. The forward telescope efficiency is 0.95. The telescope beam size varies from 56.7$''$ at 31 GHz to 35.6$''$ at 49.5 GHz. All data were analyzed
using the GILDAS package\footnote{\texttt{http://www.iram.fr/IRAMFR/GILDAS}}.

\section{Results} \label{results}

\subsection{Quantum chemical calculations of H$_2$CCCN} \label{qcc_h2cccn}

Several theoretical calculations have been done on the H$_2$CCCN radical. \citet{Fenistein1969} and \citet{Hinchliffe1977} calculated only the $C_{2v}$-symmetry structure in that two protons are equivalent relative to the linear carbon-chain backbone. Later, \citet{Mayer1998} and \citet{Parkinson1999} predicted a stable $C_s$-symmetry structure, both using density functional theory calculations, with the bending angle $\angle$CCC=164.9$^{\circ}$ and in the standard coupled cluster approach level with $\angle$CCC=149.1$^{\circ}$ \citep{Parkinson1999}. The energy barrier from the $C_s$ minimum to the $C_{2v}$ saddle point of H$_2$CCCN was calculated to be 1190 cm$^{-1}$ \citep{Parkinson1999}.

We optimized the geometry of the H$_2$CCCN radical at the spin-restricted coupled cluster method with single, double, and perturbative triple excitations (RCCSD(T)) and an explicitly correlated approximation (F12A) \citep{Adler2007,Knizia2009} with all electrons (valence and core) correlated and the Dunning's correlation consistent basis sets with polarized core-valence triple-$\zeta$ for explicitly correlated calculations (cc-pCVTZ-F12; \citealt{Hill2010a,Hill2010b}). At the optimized geometry, electric dipole moment components were calculated at the same level of theory as that for the geometrical optimization. We derive a value for $\mu_a$ of 3.6 D. The H$_2$CCCN radical in the $^{2}A'$ ground electronic state adopts a bent geometry with a $\angle$CCC angle of 147$^{\circ}$, as it can be seen in Fig. \ref{structure}. Using RCCSD(T)-F12/cc-pCVTZ-F12 level of theory we obtained the inversion barrier of $V_i$ = 356 cm$^{-1}$ at the C$_{2v}$ saddle point. These calculations were performed using the Molpro 2020 ab initio program package \citep{Werner2020}.

The fine and hyperfine coupling constants were estimated using the B3LYP hybrid density functional \citep{Becke1993} with the augmented diffuse basis set (aug-cc-pVTZ; \citealt{Woon1995}). Harmonic and anharmonic vibrational frequencies were computed using second-order M{\o}ller-Plesset perturbation (MP2; \citealt{Moller1934}) with the cc-pVTZ basis set \citep{Woon1995} level of theory to estimate the centrifugal distortion constants and the vibration–rotation interaction contribution to the rotational constants. These calculations were performed using the Gaussian16 program package \citep{Frisch2016}. The calculated molecular parameters are summarized in Table \ref{constants}.

\subsection{Rotational spectrum of H$_2$CCCN} \label{lab_h2cccn}

\begin{figure}
\centering
\includegraphics[angle=0,width=\columnwidth]{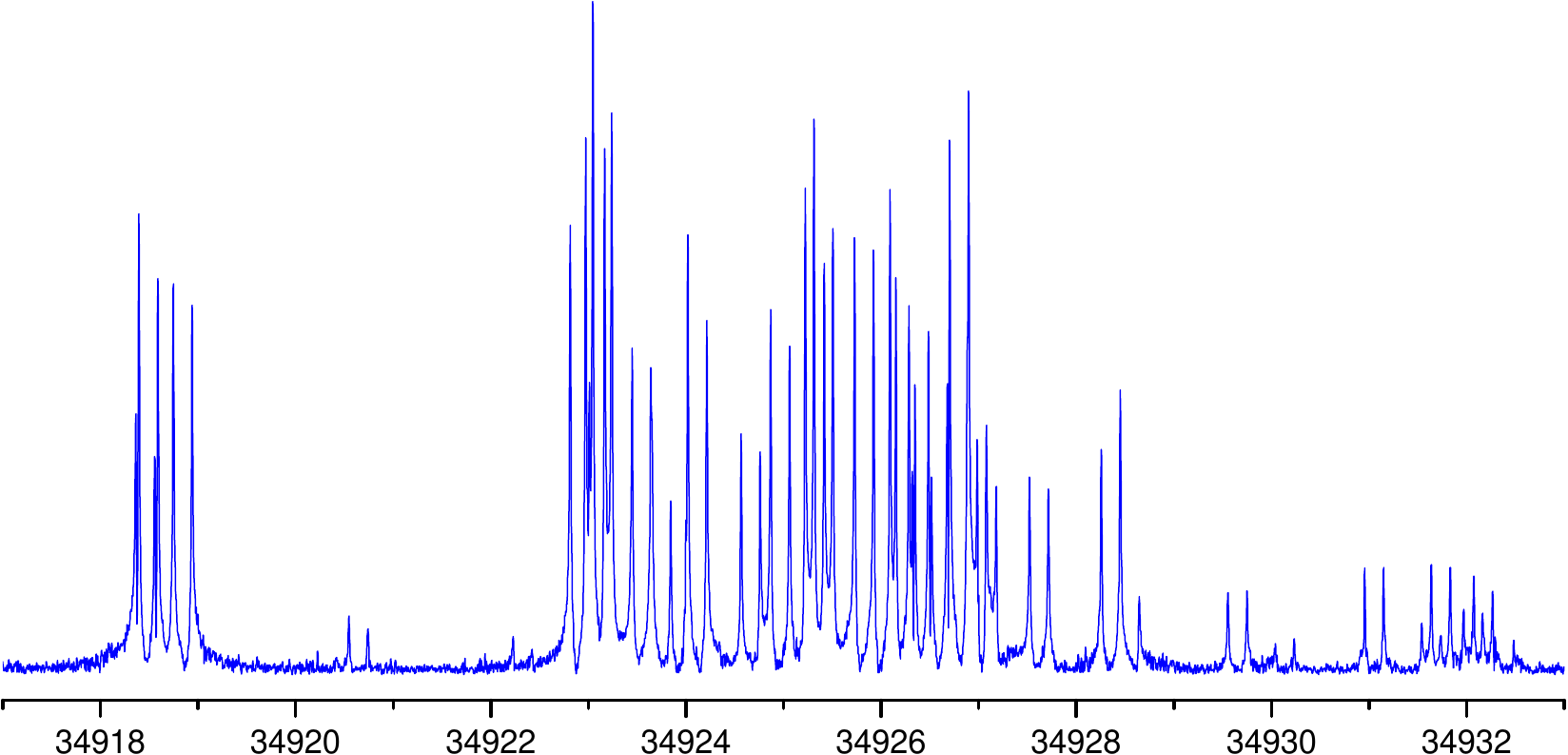}
\caption{FTMW spectrum for the H$_2$CCCN radical showing the 4$_{0,4}$-3$_{0,3}$ rotational transition at 34.9 GHz. Much weaker
additional hyperfine components, which are not shown in this figure, were also observed. The abscissa corresponds
to the frequency of the lines in MHz. The spectra were achieved by 200 shots of accumulation and a step scan of
1 MHz with a repetition rate of 10 Hz. The coaxial arrangement of the adiabatic expansion and the resonator axis
produces an instrumental Doppler doubling. The resonance frequencies are calculated as the average of the two Doppler
components.}
\label{ftmw}
\end{figure}

Due to the bent geometry for H$_2$CCCN radical in the $^{2}A'$ ground state, there are two equivalent configurations which correspond to different minima of the double minimum potential of the inversion motion. This pair of tunneling inversion states $\nu_{inv}$=$0^{\pm}$, will originate a doublet pattern in the rotational spectrum for H$_2$CCCN. Required by the Pauli's exclusion principle, ortho and para species due to two hydrogen nuclei in H$_2$CCCN correspond to the inversion states $0^+$ and $0^-$, respectively, in the even $K_a$ rotational levels. For odd $K_a$ rotational levels ortho/para changes and $0^+$ is para, while $0^-$ is ortho. The $0^+$ level is considered to be always the lower splitting sublevel.

A total of four groups of paramagnetic lines around 8.7 GHz, 17.4 GHz, 26.2 GHz, and 34.9 GHz were observed in our experiment. An example is shown in Fig. \ref{ftmw}. The H$_2$CCCN radical was readily confirmed as the spectral carrier based on the following arguments. (i) The observed transition frequencies agree well with the calculated frequencies, (ii) each transition has a hyperfine spectral structure similar to that expected for an open-shell species with three coupling nuclei, and (iii) the lines exhibit the paramagnetic behavior.

Assignment of the fine and hyperfine structure was first achieved for the spectrum of para-H$_2$CCCN ($K_a$ = 0 transitions for $0^-$ sublevel), in which the higher-$N$ transitions presented relatively simple structures. By a least-squares fitting, the spin-rotation interaction constants and the hyperfine interaction constants for the nitrogen nucleus were determined, where hyperfine interaction constants for the two hydrogen nuclei cannot be determined due to I(2H) = 0 in para-H$_2$CCCN, since this species has parity +1 for the $a$-axis rotation. Then the remaining H$_2$CCCN transition lines were assigned to the ortho-H$_2$CCCN ($K_a$ = 0 transitions for $0^+$ sublevel) spectrum. Slightly different effective molecular constants from the ones of para-H$_2$CCCN including the hyperfine interaction constants for the two hydrogen nuclei were obtained for ortho-H$_2$CCCN with I(2H) = 1, parity -1 for the $a$-axis rotation. A total of 36 and 137 hyperfine components from $K_a$=0 rotational transitions were measured for para-H$_2$CCCN and ortho-H$_2$CCCN species, respectively ( see Tables \ref{lab_freq_1} and \ref{lab_freq_2}).

All the lines for each species were independently analyzed using a standard Hamiltonian for an asymmetric top molecule with a doublet electronic state ($^2A^{'}$) and three non-zero-spin nuclei. The coupling scheme for angular momenta of rotation $\textbf{N}$, electron spin $\textbf{S}$, and nuclear spins $\textbf{I}$ of two equivalent protons and one nitrogen nucleus is
\textbf{J}\,=\,\textbf{N}\,+\,\textbf{S}, \textbf{F}$_1$\,=\,\textbf{J}\,+\,\textbf{I}$_H$, and  \textbf{F}\,=\,\textbf{F}$_1$\,+\,\textbf{I}$_N$. The molecular constants derived from the analysis are shown in Table \ref{constants}.

Although we have succeeded in assigning and analyzing the H$_2$CCCN spectra of the $K_a$ = 0 transitions by considering a tunneling inversion effect and by applying an effective Hamiltonian to the ortho/para species, information of the tunneling inversion motion, such as inversion barrier $V_i$, inversion-vibration frequency $\nu_i$, and inversion-vibration splitting $\delta$$\nu_i$$^{\pm}$, could not be obtained directly from our spectral analysis. The abnormally large inversion-vibration dependence of the $(B+C)$/2 constant indicated that there is a strong rotation-dependent perturbation in this radical.

\begin{table}
\small
\caption{Spectroscopic parameters of H$_2$CCCN (all in MHz).}
\label{constants}
\centering
\begin{tabular}{lccc}
\hline
\hline
\multicolumn{1}{c}{Parameter}  & \multicolumn{1}{c}{ortho-H$_2$CCCN } & \multicolumn{1}{c}{para-H$_2$CCCN }& \multicolumn{1}{c}{Theoretical\,$^a$} \\
\hline
$A$                         &  ~        [158105.0]\,$^b$     &  ~          [158105.0]     &  ~     158105.0    \\
$(B+C)/2$                   &  ~      4365.86049(12)\,$^c$   &  ~       4365.30705(21)    &  ~      4382.2     \\
$B-C$                       &  ~            [120.0]          &  ~           [120.0]       &  ~       120.0     \\
$\Delta_N$                  &  ~         0.0009124(48)       &  ~       0.0015943(86)     &  ~      0.00182    \\
$\Delta_{NK}$               &  ~           [$-$0.338]        &  ~           [$-$0.338]    &  ~     $-$0.338    \\
$\Delta_K$                  &  ~            [39.10]          &  ~           [39.10]       &  ~       39.10     \\
$\delta_N$                  &  ~           [0.000276]        &  ~          [0.000276]     &  ~     0.000276    \\
$\delta_K$                  &  ~            [0.430]          &  ~           [0.430]       &  ~       0.430     \\
$\varepsilon_{aa}$          &  ~            [278.0]          &  ~           [278.0]       &  ~       278.0     \\
$\varepsilon_{bb}$          &  ~            [0.489]          &  ~           [0.489]       &  ~       0.489     \\
$\varepsilon_{cc}$          &  ~        $-$12.42269(86)      &  ~       $-$12.83977(91)   &  ~      $-$10.3    \\
$a_F$$^{\rm(N)}$            &  ~          9.58207(87)        &  ~        9.50926(99)      &  ~       2.71      \\
$T_{aa}$$^{\rm(N)}$         &  ~         $-$14.4086(17)      &  ~        $-$14.3071(15)   &  ~     $-$15.8     \\
$T_{bb}$$^{\rm(N)}$         &  ~             [26.3]          &  ~            [26.3]       &  ~       26.3      \\
$\chi_{aa}$$^{\rm(N)}$      &  ~         $-$4.0506(19)       &  ~        $-$4.0606(32)    &  ~     $-$4.43     \\
$\chi_{aa}$$^{\rm(N)}$      &  ~             [2.18]          &  ~            [2.18]       &  ~        2.18     \\
$a_F$$^{\rm(H)}$            &  ~          131.627(89)        &  ~           $-$           &  ~      117.7      \\
$T_{aa}$$^{\rm(H)}$         &  ~           6.4816(16)        &  ~           $-$           &  ~       6.82      \\
$T_{bb}$$^{\rm(H)}$         &  ~           [$-$1.97]         &  ~            $-$          &  ~      $-$1.97    \\
$N_{lines}$                 &  ~              137            &  ~            36           &  ~        $-$      \\
$rms$ /kHz                  &  ~              2.9            &  ~           2.6           &  ~        $-$      \\
\hline
\end{tabular}
\tablefoot{
\tablefoottext{a}{See text for details.}
\tablefoottext{b}{Values in brackets were fixed to the theoretical values.}
\tablefoottext{c}{Numbers in parentheses are 1$\sigma$ uncertainties in units of the last digits.}
}
\end{table}

\subsection{Detection of H$_2$CCCN in TMC-1} \label{det_h2cccn}

\begin{figure}
\centering
\includegraphics[angle=0,width=\columnwidth]{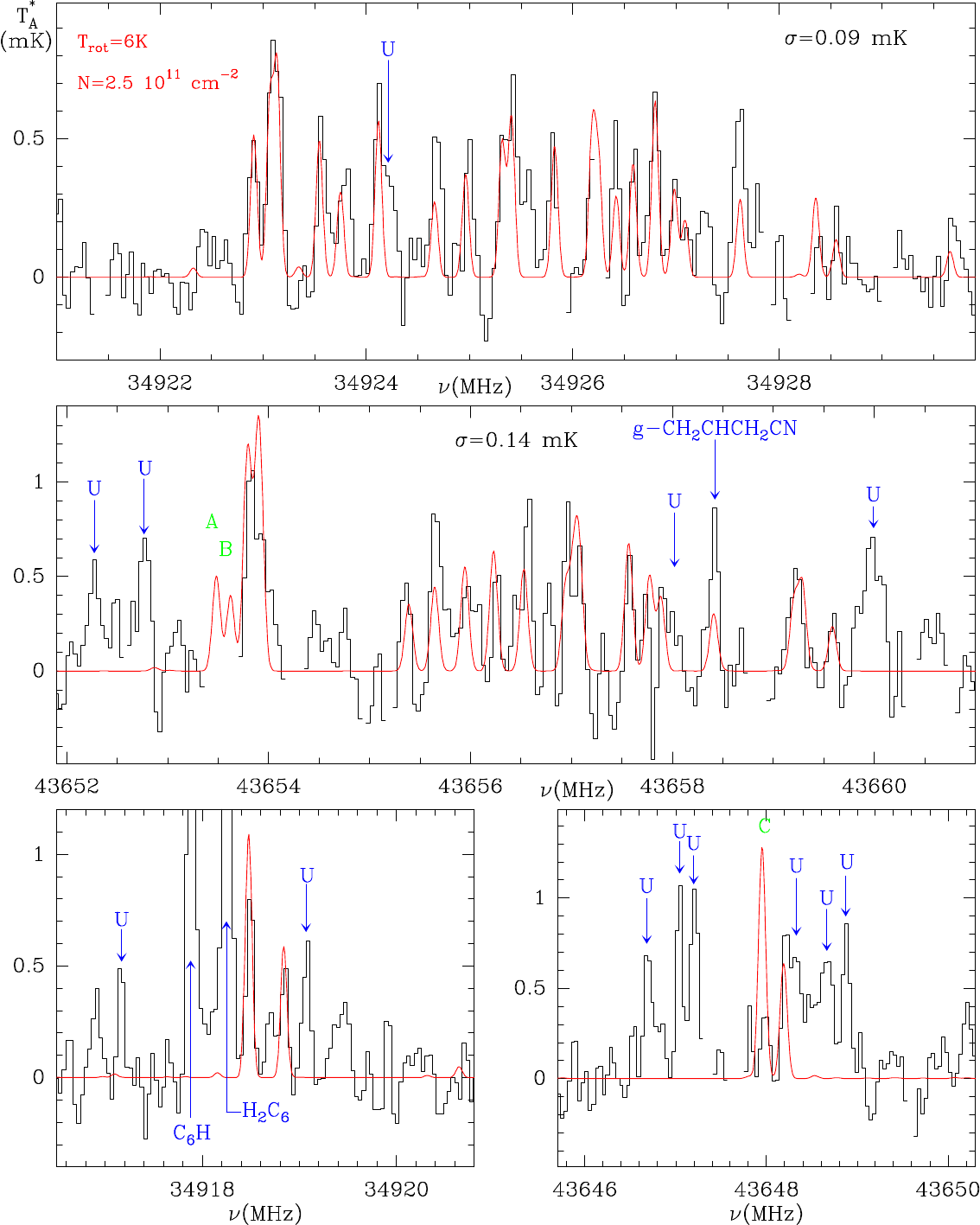}
\caption{Observed rotational transitions of H$_2$CCCN in TMC-1 in the 31.0-50.4 GHz range.
The abscissa corresponds to the rest frequency, assuming a local standard of rest velocity of 5.83\,\kms. The ordinate is antenna temperature in millikelvins. Curves shown in red are the computed synthetic spectra for T$_{rot}$=6\,K, N(ortho-H$_2$CCCN)=1.4$\times$10$^{11}$ cm$^{-2}$ and N(para-H$_2$CCCN)=1.1$\times$10$^{11}$ cm$^{-2}$.} \label{quijote}
\end{figure}

Using the molecular constants derived for ortho- and  para-H$_2$CCCN, we predicted the rotational spectra of each species. We considered the ortho and para species separately as there are no radiative or collisional transitions between them. The predictions for ortho-H$_2$CCCN include the rotational transitions with $K_a$ even for the 0$^+$ sublevel and those with $K_a$ odd for 0$^-$ sublevel, while para-H$_2$CCCN predictions contain the rotational transitions with $K_a$ even for the 0$^-$ sublevel and those with $K_a$ odd for 0$^+$ sublevel. Predictions for $K_a$=0 transitions have uncertainties smaller than 30 kHz up to $N$=10. They were implemented in the MADEX code \citep{Cernicharo2012} to compute column densities. We adopted a dipole moment of 3.60\,D as derived from our calculations.

Two rotational transitions for H$_2$CCCN with $K_a$ = 0 are covered by our QUIJOTE survey, at 34.9 and 43.6 GHz. We observed two groups of lines at the predicted frequencies for the 4$_{0,4}$-3$_{0,3}$ and 5$_{0,5}$-4$_{0,4}$ transitions. Figure \ref{quijote} shows the lines corresponding to these rotational transitions. More than 30 hyperfine components are observed in TMC-1 with a sensitivity of 0.09 mK for $N$=4 and of 0.14 mK for $N$=5. Our frequency switching observing procedure affects to some of these lines. The forest of lines for each transition covers more than 10  MHz (see Fig. \ref{quijote}) and, hence, the hyperfine components separated by 8 or 10 MHz, i.e., the frequency throws of the observations, will affect each other. In addition, at this level of sensitivity, transitions from other species and from unidentified lines could also affect the observed intensities. Some of the blends arising between the lines and their negative features at $\pm$8 MHz and $\pm$10 MHz produced in the folding of the frequency switching, have been identified and blanked in the data. As an example, the lines of C$_6$H and H$_2$C$_6$ appearing around 34918 MHz in the bottom left panel of the figure, will have negative counterparts at 34926 and 34928 MHz, which are within the frequency range of the $4_{0,4}-3_{0,3}$ transition. The corresponding channels have been cleaned in the top panel of Fig. \ref{quijote}. Three hyperfine components labelled as A, B and C in Fig. \ref{quijote} are affected by other lines of H$_2$CCCN and by unidentified lines. For other blending situations the removal of the negative feature is much more delicate and has not been performed in the final data presented in Fig. \ref{quijote}. It is worth nothing to note that each of the negative features produced by a line will have half of its intensity, which for the strongest features of Fig. \ref{quijote}, means intensities of 0.3-0.4 mK. Nevertheless, the number of detected lines is large enough to support the identification of H$_2$CCCN in TMC-1 and to ensure a reliable determination of the column density.

The observed lines of H$_2$CCCN correspond to transitions with upper energy levels of 4.2 and 6.3 K for $N$=4 and 5, respectively. Hence, the rotational temperature (T$_{rot}$) can not be derived in a reasonable way from the observations, except if it is rather low. We have explored values for T$_{rot}$ between 5 and 10 K and found that the derived column density for the two inversion states is not very sensitive to the adopted value for T$_{rot}$ \cite[see, e.g.,][]{Cernicharo2021}. In order to estimate the expected rotational temperature of the 4$_{0,4}$-3$_{0,3}$ and 5$_{0,5}$-4$_{0,4}$ transitions of H$_2$CCCN, we have analyzed the excitation conditions of a similar molecule, H$_2$CCC. For this species, the collisional rates are available \citep{Khalifa2019}, and the dipole moment is also similar to that of our molecule. For a volume density of 2-3$\times$10$^4$ cm$^{-3}$, typical of TMC-1, and assuming optically thin emission, we obtain rotational temperatures between 5.6-7\,K for the $J_u$=4 and 5 transitions of H$_2$CCC. Adopting a value of 6\,K for the two observed transitions of H$_2$CCCN, we derive a column density of (1.4$\pm$0.2)$\times$10$^{11}$ cm$^{-2}$ and of (1.1$\pm$0.2)$\times$10$^{11}$ cm$^{-2}$ for ortho-H$_2$CCCN and para-H$_2$CCCN, respectively. These column densities indicate that, within the uncertainties, both inversion states are equally populated. Adopting a rotational temperature of 7\,K introduces an increase of the column density of $\sim$10\%.

Two additional transitions for each value of $N$ corresponding to $K_a$=1 could be also within the QUIJOTE frequency coverage. However, the frequency predictions for these lines are unreliable as the $A$ rotational constant value has not been determined from the laboratory data. Moreover, using the value of the constant $A$ estimated from our calculations, the upper rotational levels of these $K_a$=1 transitions will be around 14\,K. Hence, for rotational temperatures below 10\,K the intensity of these lines is expected to be much lower than those of the transitions with $K_a$=0, and below the sensitivity of QUIJOTE.

\section{Chemistry of H$_2$CCCN} \label{discussion}

\begin{figure}
\centering
\includegraphics[angle=0,width=\columnwidth]{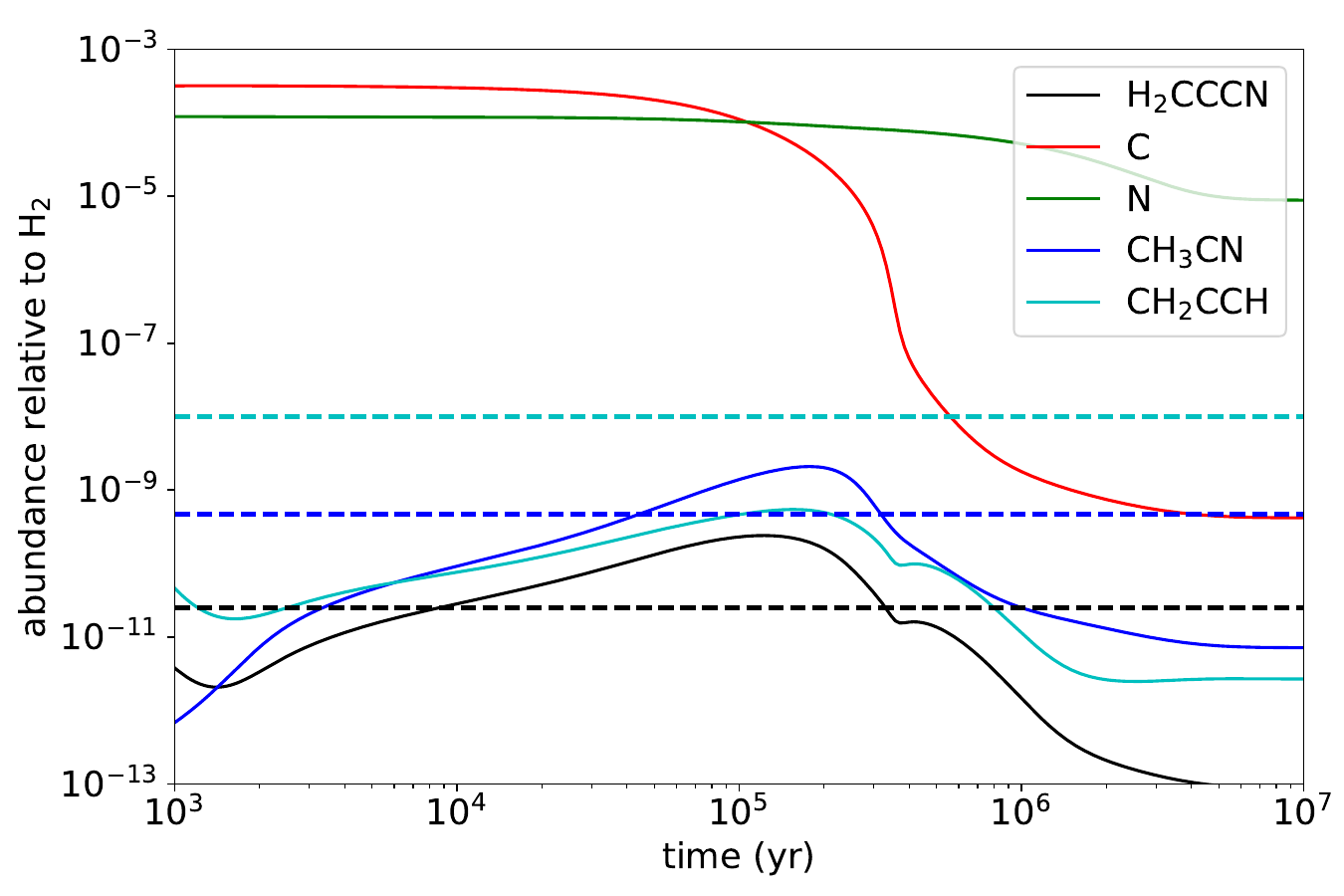}
\caption{Abundances calculated with the chemical model for H$_2$CCCN and for its potential precursors. The horizontal dashed lines correspond to the values observed in TMC-1: H$_2$CCCN (this work), CH$_3$CN \citep{Cabezas2021}, and CH$_2$CCH \citep{Agundez2022}.} \label{fig:abun}
\end{figure}

There are several chemical reactions that can form H$_2$CCCN in TMC-1, although there is little experimental or theoretical information on most of them. In fact, the chemical databases UMIST \citep{McElroy2013} and KIDA \citep{Wakelam2015} do not contain any reaction of formation of H$_2$CCCN. Among the potential routes to H$_2$CCCN we can consider the following neutral-neutral reactions which happen with H atom elimination:
\begin{equation}
\rm C + CH_3CN \rightarrow H_2CCCN + H, \label{reac:c+ch3cn}
\end{equation}
\begin{equation}
\rm CH + CH_2CN \rightarrow H_2CCCN + H, \label{reac:ch+ch2cn}
\end{equation}
\begin{equation}
\rm C_2H + H_2CN \rightarrow H_2CCCN + H, \label{reac:c2h+h2cn}
\end{equation}
\begin{equation}
\rm C_2H_3 + CN \rightarrow H_2CCCN + H. \label{reac:c2h3+cn}
\end{equation}
\begin{equation}
\rm N + CH_2CCH \rightarrow H_2CCCN + H. \label{reac:n+ch2cch}
\end{equation}
\begin{equation}
\rm NH + H_2C_3 \rightarrow H_2CCCN + H. \label{reac:n+h2c3}
\end{equation}
\begin{equation}
\rm NH_2 + C_3H \rightarrow H_2CCCN + H. \label{reac:n+h2c3}
\end{equation}

To evaluate whether these reactions can produce H$_2$CCCN with an abundance similar to that observed we plugged them into a chemical model. We used as starting point the chemical model built by \cite{Cabezas2021} to study the chemistry of CH$_2$CCCN, which uses the chemical network RATE12 from the UMIST database \citep{McElroy2013}, with updates from \cite{Loison2014} and \cite{Marcelino2021}. The physical parameters assumed are the typical ones of cold dark clouds, i.e., a density of H nuclei of 2\,$\times$\,10$^4$ cm$^{-3}$, a gas kinetic temperature of 10 K, a cosmic-ray ionization rate of H$_2$ of 1.3\,$\times$\,10$^{-17}$ s$^{-1}$, a visual extinction of 30 mag, and the set of low-metal elemental abundances (see, e.g., \citealt{Agundez2013}). We added H$_2$CCCN as new species and included the above seven neutral-neutral reactions with a rate coefficient of 10$^{-10}$ cm$^3$ s$^{-1}$, which is typical of neutral-neutral reactions that are fast at low temperature. We considered that H$_2$CCCN is destroyed by reacting with neutral atoms, such as H, O, N, and C, with a rate coefficient of 10$^{-10}$ cm$^3$ s$^{-1}$, and with abundant positive cations, such as H$^+$ and C$^+$, with a rate coefficient of 10$^{-9}$ cm$^3$ s$^{-1}$ at 300 K and a dependence with temperature of the type $T^{-0.5}$.

The results are shown in Fig.~\ref{fig:abun}. The first conclusion is that H$_2$CCCN is indeed produced with a peak abundance of the order of the observed value at a time of a few 10$^5$ yr. The second conclusion is that the main reactions of formation of H$_2$CCCN are by far reactions~(\ref{reac:c+ch3cn}) and (\ref{reac:n+ch2cch}), because they involve as reactants atomic carbon and atomic nitrogen, respectively, which are very abundant (see Fig.~\ref{fig:abun}). Interestingly, reaction~(\ref{reac:c+ch3cn}) has been recently studied by \cite{Hickson2021}. These authors measured the rate coefficient to be (3-4)\,$\times$\,10$^{-10}$ cm$^3$ s$^{-1}$ in the temperature range 50-296 K and the H atom yield to be 0.63 at 177 K. They also carried out quantum chemical calculations, which point to H$_2$CCCN and H$_2$CCNC as the most likely products on the triplet potential energy surface. Reaction~(\ref{reac:n+ch2cch}) has not been studied to our knowledge. \cite{Loison2017} suggest that the reaction is fast, although they favor C$_2$H$_2$ + HCN and HC$_3$N + H$_2$ as products. It would be interesting to investigate whether this reaction can produce H$_2$CCCN at low temperature. In summary, the reaction C + CH$_3$CN, and perhaps also the reaction N + CH$_2$CCH, emerge as the most likely routes to H$_2$CCCN. The species H$_2$CCNC is also predicted to be formed in the C + CH$_3$CN reaction and thus it is an interesting target to be searched for in TMC-1.

Alternatively, the cyanovinyl radical could also be formed in the dissociative recombination with electrons of positive ions such as H$_3$C$_3$N$^+$ or H$_4$C$_3$N$^+$. The astrochemical databases UMIST and KIDA consider that the dissociative recombination of H$_4$C$_3$N$^+$ produces vinyl cyanide (CH$_2$CHCN). However, the product distribution of this reaction has not been measured and H$_2$CCCN could also be formed.

\section{Conclusions}

We report the detection in TMC-1 of the radical H$_2$CCCN based on accurate laboratory spectroscopy of the $N$=1, 2, 3 and 4 $K_a$=0 rotational transitions of this species, and on the QUIJOTE line survey of this prestellar cold core. We have identified 30 hyperfine components in TMC-1. Chemical modeling indicate that the reaction C + CH$_3$CN, and possibly N + CH$_2$CCH, as the most likely paths to H$_2$CCCN in TMC-1.

\begin{acknowledgements}
The present study was supported by ERC through the grant ERC-2013-Syg-610256-NANOCOSMOS, Ministry of Science and Technology of Taiwan through project MOST 104-2113-M-009-202, JSPS KAKENHI through Grant 06640644, and Ministerio de Ciencia e Innovaci\'on of Spain through projects PID2019-106110GB-I00, PID2019-107115GB-C21 / AEI / 10.13039/501100011033, and PID2019-106235GB-I00. C.C., M.A, Y. E and J. C. thank Ministry of Science and Technology of Taiwan and Consejo Superior de Investigaciones Cient\'ificas for funding support under the MOST-CSIC Mobility Action 2021 (Grant 11-2927-I-A49-502 and OSTW200006). J. T. thanks the JSPS Postdoctoral Fellowship for Foreign Researchers and Grant-in-Aid for JSPS Fellows.

\end{acknowledgements}

\normalsize

\begin{appendix}

\section{Laboratory transition frequencies for H$_2$CCCN}
\label{lab_freq}
The laboratory measurements of H$_2$CCCN described in Sect.\,\ref{labo_observations} have permitted to measure
137 and 36 hyperfine components for ortho-H$_2$CCCN and para-H$_2$CCCN, respectively. The observed
frequencies and quantum number assignments are given in Tables \ref{lab_freq_1} and \ref{lab_freq_2}. A file with predictions up to $N$=30 has been uploaded to the CDS.

\onecolumn
\begin{tiny}
\begin{longtable}{cccccccccccccc}
\caption[]{Observed transition frequencies for ortho-H$_2$CCCN.
\label{lab_freq_1}}\\
\hline
\hline
 $N'$ & $K'_a$ & $K'_c$ & $J'$ & $F'_1$ & $F'$ & $N''$ & $K''_a$ & $K''_c$ & $J''$ & $F''_1$ & $F''$ & $\nu_{obs}$  &   Obs-Calc \\
      &        &        &      &        &        &      &        &      &        &        &      &   (MHz)      &   (MHz)   \\
\hline
\endfirsthead
\caption{continued.}\\
\hline
\hline
 $N'$ & $K'_a$ & $K'_c$ & $J'$ & $F'_1$ & $F'$ & $N''$ & $K''_a$ & $K''_c$ & $J''$ & $F''_1$ & $F''$ & $\nu_{obs}$  &   Obs-Calc \\
      &        &        &      &        &        &    &        &      &        &        &      &   (MHz)      &   (MHz)      \\
\hline
\endhead
\hline
\endfoot
\hline
\endlastfoot
\hline
  1 & 0 & 1 & 1.5 & 2.5 & 2.5 &  0 & 0 & 0 & 0.5 & 1.5 & 2.5 &   8720.316   &  0.000  \\
  1 & 0 & 1 & 1.5 & 0.5 & 0.5 &  0 & 0 & 0 & 0.5 & 0.5 & 0.5 &   8722.196   & -0.002  \\
  1 & 0 & 1 & 1.5 & 2.5 & 1.5 &  0 & 0 & 0 & 0.5 & 1.5 & 1.5 &   8722.801   & -0.000  \\
  1 & 0 & 1 & 0.5 & 0.5 & 1.5 &  0 & 0 & 0 & 0.5 & 1.5 & 2.5 &   8724.701   &  0.000  \\
  1 & 0 & 1 & 1.5 & 0.5 & 0.5 &  0 & 0 & 0 & 0.5 & 0.5 & 1.5 &   8727.134   & -0.003  \\
  1 & 0 & 1 & 1.5 & 2.5 & 1.5 &  0 & 0 & 0 & 0.5 & 1.5 & 0.5 &   8727.742   & -0.001  \\
  1 & 0 & 1 & 1.5 & 2.5 & 2.5 &  0 & 0 & 0 & 0.5 & 1.5 & 1.5 &   8728.041   & -0.000  \\
  1 & 0 & 1 & 1.5 & 2.5 & 3.5 &  0 & 0 & 0 & 0.5 & 1.5 & 2.5 &   8729.730   &  0.000  \\
  1 & 0 & 1 & 1.5 & 1.5 & 1.5 &  0 & 0 & 0 & 0.5 & 0.5 & 0.5 &   8731.345   &  0.001  \\
  1 & 0 & 1 & 0.5 & 1.5 & 2.5 &  0 & 0 & 0 & 0.5 & 1.5 & 2.5 &   8731.598   &  0.002  \\
  1 & 0 & 1 & 1.5 & 0.5 & 1.5 &  0 & 0 & 0 & 0.5 & 0.5 & 1.5 &   8731.743   & -0.001  \\
  1 & 0 & 1 & 0.5 & 0.5 & 1.5 &  0 & 0 & 0 & 0.5 & 1.5 & 1.5 &   8732.429   &  0.003  \\
  1 & 0 & 1 & 1.5 & 1.5 & 2.5 &  0 & 0 & 0 & 0.5 & 0.5 & 1.5 &   8732.533   &  0.001  \\
  1 & 0 & 1 & 0.5 & 0.5 & 0.5 &  0 & 0 & 0 & 0.5 & 1.5 & 0.5 &   8732.721   &  0.006  \\
  1 & 0 & 1 & 1.5 & 1.5 & 0.5 &  0 & 0 & 0 & 0.5 & 0.5 & 0.5 &   8732.986   & -0.001  \\
  1 & 0 & 1 & 0.5 & 0.5 & 1.5 &  0 & 0 & 0 & 0.5 & 1.5 & 0.5 &   8737.370   &  0.003  \\
  1 & 0 & 1 & 1.5 & 1.5 & 0.5 &  0 & 0 & 0 & 0.5 & 0.5 & 1.5 &   8737.925   & -0.001  \\
  1 & 0 & 1 & 0.5 & 1.5 & 2.5 &  0 & 0 & 0 & 0.5 & 1.5 & 1.5 &   8739.324   &  0.002  \\
  1 & 0 & 1 & 0.5 & 1.5 & 1.5 &  0 & 0 & 0 & 0.5 & 1.5 & 2.5 &   8740.312   & -0.001  \\
  1 & 0 & 1 & 0.5 & 1.5 & 0.5 &  0 & 0 & 0 & 0.5 & 1.5 & 1.5 &   8740.999   & -0.003  \\
  1 & 0 & 1 & 0.5 & 1.5 & 1.5 &  0 & 0 & 0 & 0.5 & 1.5 & 1.5 &   8748.038   & -0.001  \\
  2 & 0 & 2 & 1.5 & 2.5 & 2.5 &  1 & 0 & 1 & 0.5 & 1.5 & 1.5 &  17449.833   &  0.003  \\
  2 & 0 & 2 & 2.5 & 3.5 & 3.5 &  1 & 0 & 1 & 1.5 & 2.5 & 3.5 &  17450.968   & -0.001  \\
  2 & 0 & 2 & 2.5 & 1.5 & 0.5 &  1 & 0 & 1 & 1.5 & 1.5 & 0.5 &  17452.212   & -0.003  \\
  2 & 0 & 2 & 2.5 & 3.5 & 2.5 &  1 & 0 & 1 & 1.5 & 2.5 & 2.5 &  17454.796   & -0.001  \\
  2 & 0 & 2 & 1.5 & 1.5 & 2.5 &  1 & 0 & 1 & 0.5 & 1.5 & 1.5 &  17457.567   &  0.002  \\
  2 & 0 & 2 & 2.5 & 1.5 & 0.5 &  1 & 0 & 1 & 1.5 & 0.5 & 1.5 &  17458.390   & -0.007  \\
  2 & 0 & 2 & 2.5 & 1.5 & 1.5 &  1 & 0 & 1 & 1.5 & 0.5 & 1.5 &  17459.322   & -0.000  \\
  2 & 0 & 2 & 2.5 & 3.5 & 2.5 &  1 & 0 & 1 & 1.5 & 2.5 & 1.5 &  17460.035   & -0.002  \\
  2 & 0 & 2 & 2.5 & 3.5 & 3.5 &  1 & 0 & 1 & 1.5 & 2.5 & 2.5 &  17460.383   & -0.000  \\
  2 & 0 & 2 & 2.5 & 3.5 & 4.5 &  1 & 0 & 1 & 1.5 & 2.5 & 3.5 &  17460.787   &  0.002  \\
  2 & 0 & 2 & 1.5 & 0.5 & 1.5 &  1 & 0 & 1 & 0.5 & 0.5 & 1.5 &  17462.332   &  0.001  \\
  2 & 0 & 2 & 2.5 & 1.5 & 2.5 &  1 & 0 & 1 & 1.5 & 1.5 & 2.5 &  17462.511   & -0.001  \\
  2 & 0 & 2 & 2.5 & 1.5 & 0.5 &  1 & 0 & 1 & 1.5 & 0.5 & 0.5 &  17463.000   & -0.003  \\
  2 & 0 & 2 & 2.5 & 2.5 & 2.5 &  1 & 0 & 1 & 1.5 & 1.5 & 1.5 &  17463.285\,$^a$  & -0.001  \\
  2 & 0 & 2 & 2.5 & 2.5 & 3.5 &  1 & 0 & 1 & 0.5 & 1.5 & 2.5 &  17463.285\,$^a$  & -0.002  \\
  2 & 0 & 2 & 2.5 & 1.5 & 2.5 &  1 & 0 & 1 & 1.5 & 0.5 & 1.5 &  17463.306   &  0.005  \\
  2 & 0 & 2 & 1.5 & 1.5 & 1.5 &  1 & 0 & 1 & 0.5 & 1.5 & 1.5 &  17463.433   &  0.006  \\
  2 & 0 & 2 & 2.5 & 2.5 & 1.5 &  1 & 0 & 1 & 1.5 & 1.5 & 0.5 &  17463.518   & -0.001  \\
  2 & 0 & 2 & 2.5 & 1.5 & 1.5 &  1 & 0 & 1 & 1.5 & 0.5 & 0.5 &  17463.928   & -0.001  \\
  2 & 0 & 2 & 1.5 & 2.5 & 3.5 &  1 & 0 & 1 & 1.5 & 1.5 & 2.5 &  17464.242   &  0.001  \\
  2 & 0 & 2 & 2.5 & 2.5 & 3.5 &  1 & 0 & 1 & 1.5 & 2.5 & 3.5 &  17465.156   &  0.002  \\
  2 & 0 & 2 & 1.5 & 2.5 & 2.5 &  1 & 0 & 1 & 0.5 & 0.5 & 1.5 &  17465.442   & -0.001  \\
  2 & 0 & 2 & 1.5 & 1.5 & 2.5 &  1 & 0 & 1 & 0.5 & 1.5 & 2.5 &  17466.281   & -0.001  \\
  2 & 0 & 2 & 1.5 & 0.5 & 1.5 &  1 & 0 & 1 & 0.5 & 0.5 & 0.5 &  17466.983   & -0.000  \\
  2 & 0 & 2 & 1.5 & 0.5 & 0.5 &  1 & 0 & 1 & 0.5 & 1.5 & 1.5 &  17467.473   &  0.001  \\
  2 & 0 & 2 & 2.5 & 2.5 & 2.5 &  1 & 0 & 1 & 1.5 & 0.5 & 1.5 &  17467.828   &  0.003  \\
  2 & 0 & 2 & 1.5 & 1.5 & 2.5 &  1 & 0 & 1 & 1.5 & 2.5 & 3.5 &  17468.148   & -0.001  \\
  2 & 0 & 2 & 1.5 & 2.5 & 1.5 &  1 & 0 & 1 & 0.5 & 0.5 & 1.5 &  17468.237   & -0.004  \\
  2 & 0 & 2 & 2.5 & 2.5 & 1.5 &  1 & 0 & 1 & 1.5 & 0.5 & 1.5 &  17469.699   & -0.002  \\
  2 & 0 & 2 & 1.5 & 2.5 & 2.5 &  1 & 0 & 1 & 1.5 & 2.5 & 2.5 &  17469.832   &  0.004  \\
  2 & 0 & 2 & 1.5 & 1.5 & 0.5 &  1 & 0 & 1 & 0.5 & 0.5 & 1.5 &  17471.760   & -0.000  \\
  2 & 0 & 2 & 1.5 & 1.5 & 1.5 &  1 & 0 & 1 & 0.5 & 1.5 & 2.5 &  17472.143   & -0.001  \\
  2 & 0 & 2 & 1.5 & 1.5 & 2.5 &  1 & 0 & 1 & 0.5 & 0.5 & 1.5 &  17473.179   &  0.001  \\
  2 & 0 & 2 & 2.5 & 2.5 & 3.5 &  1 & 0 & 1 & 1.5 & 2.5 & 2.5 &  17474.567   & -0.001  \\
  2 & 0 & 2 & 1.5 & 1.5 & 0.5 &  1 & 0 & 1 & 0.5 & 0.5 & 0.5 &  17476.408   & -0.005  \\
  2 & 0 & 2 & 1.5 & 1.5 & 2.5 &  1 & 0 & 1 & 1.5 & 2.5 & 2.5 &  17477.563   &  0.000  \\
  2 & 0 & 2 & 1.5 & 2.5 & 1.5 &  1 & 0 & 1 & 1.5 & 2.5 & 1.5 &  17477.869   &  0.003  \\
  2 & 0 & 2 & 1.5 & 1.5 & 1.5 &  1 & 0 & 1 & 0.5 & 0.5 & 1.5 &  17479.044   &  0.004  \\
  2 & 0 & 2 & 1.5 & 0.5 & 0.5 &  1 & 0 & 1 & 0.5 & 0.5 & 1.5 &  17483.086   &  0.001  \\
  2 & 0 & 2 & 1.5 & 1.5 & 1.5 &  1 & 0 & 1 & 1.5 & 2.5 & 2.5 &  17483.427   &  0.002  \\
  3 & 0 & 3 & 3.5 & 4.5 & 4.5 &  2 & 0 & 2 & 2.5 & 3.5 & 4.5 &  26182.068   &  0.000  \\
  3 & 0 & 3 & 3.5 & 2.5 & 1.5 &  2 & 0 & 2 & 2.5 & 2.5 & 1.5 &  26182.366   & -0.002  \\
  3 & 0 & 3 & 3.5 & 4.5 & 3.5 &  2 & 0 & 2 & 2.5 & 3.5 & 3.5 &  26186.066   & -0.001  \\
  3 & 0 & 3 & 3.5 & 3.5 & 3.5 &  2 & 0 & 2 & 1.5 & 1.5 & 2.5 &  26187.413   &  0.001  \\
  3 & 0 & 3 & 3.5 & 2.5 & 3.5 &  2 & 0 & 2 & 2.5 & 2.5 & 2.5 &  26189.880   &  0.000  \\
  3 & 0 & 3 & 3.5 & 2.5 & 2.5 &  2 & 0 & 2 & 2.5 & 1.5 & 2.5 &  26190.085   & -0.000  \\
  3 & 0 & 3 & 3.5 & 4.5 & 3.5 &  2 & 0 & 2 & 2.5 & 3.5 & 2.5 &  26191.654   &  0.001  \\
  3 & 0 & 3 & 3.5 & 4.5 & 4.5 &  2 & 0 & 2 & 2.5 & 3.5 & 3.5 &  26191.885   &  0.001  \\
  3 & 0 & 3 & 3.5 & 4.5 & 5.5 &  2 & 0 & 2 & 2.5 & 3.5 & 4.5 &  26192.048   &  0.006  \\
  3 & 0 & 3 & 3.5 & 2.5 & 1.5 &  2 & 0 & 2 & 2.5 & 1.5 & 1.5 &  26192.746   & -0.000  \\
  3 & 0 & 3 & 3.5 & 2.5 & 1.5 &  2 & 0 & 2 & 2.5 & 1.5 & 0.5 &  26193.672   &  0.000  \\
  3 & 0 & 3 & 2.5 & 2.5 & 2.5 &  2 & 0 & 2 & 1.5 & 1.5 & 1.5 &  26193.863   & -0.002  \\
  3 & 0 & 3 & 3.5 & 2.5 & 2.5 &  2 & 0 & 2 & 2.5 & 1.5 & 1.5 &  26194.065   &  0.001  \\
  3 & 0 & 3 & 3.5 & 3.5 & 4.5 &  2 & 0 & 2 & 2.5 & 2.5 & 3.5 &  26194.337   &  0.002  \\
  3 & 0 & 3 & 3.5 & 2.5 & 3.5 &  2 & 0 & 2 & 2.5 & 1.5 & 2.5 &  26194.408\,$^a$  &  0.004  \\
  3 & 0 & 3 & 2.5 & 2.5 & 3.5 &  2 & 0 & 2 & 1.5 & 1.5 & 2.5 &  26194.408\,$^a$  & -0.003  \\
  3 & 0 & 3 & 2.5 & 3.5 & 2.5 &  2 & 0 & 2 & 1.5 & 2.5 & 1.5 &  26194.869   & -0.000  \\
  3 & 0 & 3 & 2.5 & 3.5 & 3.5 &  2 & 0 & 2 & 2.5 & 2.5 & 2.5 &  26194.880   &  0.000  \\
  3 & 0 & 3 & 3.5 & 3.5 & 2.5 &  2 & 0 & 2 & 2.5 & 2.5 & 1.5 &  26195.096   & -0.001  \\
  3 & 0 & 3 & 3.5 & 3.5 & 3.5 &  2 & 0 & 2 & 1.5 & 2.5 & 2.5 &  26195.146   & -0.001  \\
  3 & 0 & 3 & 2.5 & 3.5 & 4.5 &  2 & 0 & 2 & 1.5 & 2.5 & 3.5 &  26195.668   &  0.001  \\
  3 & 0 & 3 & 2.5 & 2.5 & 1.5 &  2 & 0 & 2 & 1.5 & 1.5 & 0.5 &  26196.814   & -0.001  \\
  3 & 0 & 3 & 3.5 & 3.5 & 2.5 &  2 & 0 & 2 & 2.5 & 2.5 & 2.5 &  26196.965   & -0.008  \\
  3 & 0 & 3 & 2.5 & 1.5 & 2.5 &  2 & 0 & 2 & 1.5 & 0.5 & 1.5 &  26197.342   & -0.002  \\
  3 & 0 & 3 & 2.5 & 2.5 & 3.5 &  2 & 0 & 2 & 2.5 & 2.5 & 3.5 &  26197.405   & -0.001  \\
  3 & 0 & 3 & 2.5 & 3.5 & 2.5 &  2 & 0 & 2 & 1.5 & 2.5 & 2.5 &  26197.665   & -0.002  \\
  3 & 0 & 3 & 2.5 & 1.5 & 0.5 &  2 & 0 & 2 & 1.5 & 0.5 & 0.5 &  26197.994   &  0.001  \\
  3 & 0 & 3 & 3.5 & 3.5 & 4.5 &  2 & 0 & 2 & 2.5 & 3.5 & 4.5 &  26198.704   &  0.001  \\
  3 & 0 & 3 & 2.5 & 1.5 & 1.5 &  2 & 0 & 2 & 1.5 & 1.5 & 1.5 &  26198.952   &  0.001  \\
  3 & 0 & 3 & 2.5 & 3.5 & 3.5 &  2 & 0 & 2 & 2.5 & 1.5 & 2.5 &  26199.405   &  0.001  \\
  3 & 0 & 3 & 2.5 & 2.5 & 2.5 &  2 & 0 & 2 & 1.5 & 1.5 & 2.5 &  26199.727   & -0.000  \\
  3 & 0 & 3 & 2.5 & 2.5 & 1.5 &  2 & 0 & 2 & 1.5 & 2.5 & 1.5 &  26200.337   &  0.003  \\
  3 & 0 & 3 & 2.5 & 3.5 & 2.5 &  2 & 0 & 2 & 1.5 & 0.5 & 1.5 &  26200.775   & -0.005  \\
  3 & 0 & 3 & 3.5 & 3.5 & 2.5 &  2 & 0 & 2 & 2.5 & 1.5 & 2.5 &  26201.495   & -0.002  \\
  3 & 0 & 3 & 2.5 & 1.5 & 0.5 &  2 & 0 & 2 & 1.5 & 1.5 & 1.5 &  26202.045   &  0.007  \\
  3 & 0 & 3 & 2.5 & 2.5 & 3.5 &  2 & 0 & 2 & 1.5 & 2.5 & 2.5 &  26202.144   & -0.002  \\
  3 & 0 & 3 & 2.5 & 2.5 & 2.5 &  2 & 0 & 2 & 2.5 & 2.5 & 3.5 &  26202.713   & -0.010  \\
  3 & 0 & 3 & 3.5 & 3.5 & 3.5 &  2 & 0 & 2 & 2.5 & 3.5 & 3.5 &  26204.593   &  0.001  \\
  3 & 0 & 3 & 2.5 & 1.5 & 1.5 &  2 & 0 & 2 & 1.5 & 1.5 & 2.5 &  26204.813   & -0.000  \\
  3 & 0 & 3 & 2.5 & 2.5 & 1.5 &  2 & 0 & 2 & 1.5 & 0.5 & 1.5 &  26206.241   & -0.003  \\
  3 & 0 & 3 & 2.5 & 1.5 & 2.5 &  2 & 0 & 2 & 2.5 & 3.5 & 2.5 &  26209.269   &  0.006  \\
  3 & 0 & 3 & 2.5 & 2.5 & 3.5 &  2 & 0 & 2 & 2.5 & 3.5 & 3.5 &  26211.591   &  0.001  \\
  3 & 0 & 3 & 2.5 & 3.5 & 2.5 &  2 & 0 & 2 & 2.5 & 3.5 & 2.5 &  26212.700   &  0.002  \\
  4 & 0 & 4 & 4.5 & 3.5 & 4.5 &  3 & 0 & 3 & 2.5 & 3.5 & 3.5 &  34920.322   &  0.006  \\
  4 & 0 & 4 & 4.5 & 3.5 & 3.5 &  3 & 0 & 3 & 3.5 & 2.5 & 3.5 &  34920.644   & -0.001  \\
  4 & 0 & 4 & 4.5 & 3.5 & 4.5 &  3 & 0 & 3 & 2.5 & 3.5 & 4.5 &  34922.326   &  0.002  \\
  4 & 0 & 4 & 4.5 & 5.5 & 4.5 &  3 & 0 & 3 & 3.5 & 4.5 & 3.5 &  34922.914   &  0.004  \\
  4 & 0 & 4 & 4.5 & 5.5 & 5.5 &  3 & 0 & 3 & 3.5 & 4.5 & 4.5 &  34923.071   &  0.006  \\
  4 & 0 & 4 & 4.5 & 5.5 & 6.5 &  3 & 0 & 3 & 3.5 & 4.5 & 5.5 &  34923.144   &  0.008  \\
  4 & 0 & 4 & 3.5 & 4.5 & 3.5 &  3 & 0 & 3 & 2.5 & 3.5 & 2.5 &  34923.546   & -0.004  \\
  4 & 0 & 4 & 4.5 & 3.5 & 2.5 &  3 & 0 & 3 & 3.5 & 2.5 & 1.5 &  34924.664   &  0.002  \\
  4 & 0 & 4 & 4.5 & 3.5 & 3.5 &  3 & 0 & 3 & 3.5 & 2.5 & 2.5 &  34924.966   &  0.002  \\
  4 & 0 & 4 & 4.5 & 3.5 & 4.5 &  3 & 0 & 3 & 3.5 & 2.5 & 3.5 &  34925.321   &  0.005  \\
  4 & 0 & 4 & 4.5 & 4.5 & 5.5 &  3 & 0 & 3 & 3.5 & 3.5 & 4.5 &  34925.410   &  0.002  \\
  4 & 0 & 4 & 4.5 & 4.5 & 4.5 &  3 & 0 & 3 & 3.5 & 3.5 & 3.5 &  34925.826   & -0.001  \\
  4 & 0 & 4 & 3.5 & 4.5 & 4.5 &  3 & 0 & 3 & 2.5 & 3.5 & 3.5 &  34926.190   &  0.000  \\
  4 & 0 & 4 & 4.5 & 4.5 & 3.5 &  3 & 0 & 3 & 3.5 & 3.5 & 2.5 &  34926.250   & -0.000  \\
  4 & 0 & 4 & 3.5 & 3.5 & 3.5 &  3 & 0 & 3 & 2.5 & 2.5 & 2.5 &  34926.419   & -0.003  \\
  4 & 0 & 4 & 3.5 & 3.5 & 4.5 &  3 & 0 & 3 & 2.5 & 2.5 & 3.5 &  34926.582   & -0.004  \\
  4 & 0 & 4 & 3.5 & 4.5 & 5.5 &  3 & 0 & 3 & 2.5 & 3.5 & 4.5 &  34926.801   &  0.002  \\
  4 & 0 & 4 & 3.5 & 4.5 & 3.5 &  3 & 0 & 3 & 2.5 & 1.5 & 2.5 &  34926.989   &  0.003  \\
  4 & 0 & 4 & 3.5 & 2.5 & 2.5 &  3 & 0 & 3 & 2.5 & 1.5 & 1.5 &  34927.082   & -0.005  \\
  4 & 0 & 4 & 3.5 & 2.5 & 3.5 &  3 & 0 & 3 & 2.5 & 3.5 & 2.5 &  34927.619   & -0.001  \\
  4 & 0 & 4 & 3.5 & 3.5 & 2.5 &  3 & 0 & 3 & 2.5 & 2.5 & 1.5 &  34928.355   & -0.001  \\
  4 & 0 & 4 & 3.5 & 2.5 & 1.5 &  3 & 0 & 3 & 2.5 & 1.5 & 0.5 &  34928.549   & -0.000  \\
  4 & 0 & 4 & 3.5 & 3.5 & 4.5 &  3 & 0 & 3 & 3.5 & 3.5 & 4.5 &  34929.653   & -0.004  \\
  4 & 0 & 4 & 3.5 & 2.5 & 3.5 &  3 & 0 & 3 & 3.5 & 3.5 & 3.5 &  34930.138   & -0.003  \\
  4 & 0 & 4 & 3.5 & 2.5 & 3.5 &  3 & 0 & 3 & 2.5 & 1.5 & 2.5 &  34931.053   & -0.003  \\
  4 & 0 & 4 & 3.5 & 2.5 & 1.5 &  3 & 0 & 3 & 2.5 & 1.5 & 1.5 &  34931.638   &  0.002  \\
  4 & 0 & 4 & 3.5 & 3.5 & 3.5 &  3 & 0 & 3 & 2.5 & 2.5 & 3.5 &  34931.736   & -0.003  \\
  4 & 0 & 4 & 4.5 & 4.5 & 5.5 &  3 & 0 & 3 & 3.5 & 4.5 & 5.5 &  34932.067   & -0.003  \\
  4 & 0 & 4 & 3.5 & 2.5 & 2.5 &  3 & 0 & 3 & 2.5 & 2.5 & 2.5 &  34932.170   & -0.003  \\
  4 & 0 & 4 & 3.5 & 3.5 & 4.5 &  3 & 0 & 3 & 3.5 & 3.5 & 3.5 &  34933.581   & -0.004  \\
  4 & 0 & 4 & 3.5 & 3.5 & 2.5 &  3 & 0 & 3 & 2.5 & 3.5 & 2.5 &  34933.823   &  0.002  \\
  4 & 0 & 4 & 3.5 & 3.5 & 3.5 &  3 & 0 & 3 & 2.5 & 3.5 & 2.5 &  34936.215   & -0.002  \\
  4 & 0 & 4 & 4.5 & 4.5 & 4.5 &  3 & 0 & 3 & 3.5 & 4.5 & 4.5 &  34938.534   & -0.000  \\
\hline
\end{longtable}
\tablefoot{
\tablefoottext{a}{The weight of 0.5 is given in the least-squares analysis, because of overlapping with other line.}
}
\end{tiny}
\twocolumn

\onecolumn
\begin{tiny}
\begin{longtable}{cccccccccccccc}
\caption[]{Observed transition frequencies for para-H$_2$CCCN.
\label{lab_freq_2}}\\
\hline
\hline
 $N'$ & $K'_a$ & $K'_c$ & $J'$ & $F'_1$ & $F'$ & $N''$ & $K''_a$ & $K''_c$ & $J''$ & $F''_1$ & $F''$ & $\nu_{obs}$  &   Obs-Calc \\
      &        &        &      &        &        &      &        &      &        &        &      &   (MHz)      &   (MHz)   \\
\hline
\endfirsthead
\caption{continued.}\\
\hline
\hline
 $N'$ & $K'_a$ & $K'_c$ & $J'$ & $F'_1$ & $F'$ & $N''$ & $K''_a$ & $K''_c$ & $J''$ & $F''_1$ & $F''$ & $\nu_{obs}$  &   Obs-Calc \\
      &        &        &      &        &        &    &        &      &        &        &      &   (MHz)      &   (MHz)      \\
\hline
\endhead
\hline
\endfoot
\hline
\endlastfoot
\hline
1 & 0 & 1 & 1.5 & 1.5 & 1.5 &  0 & 0 & 0 & 0.5 & 0.5 & 1.5 &   8716.165  & -0.010  \\
1 & 0 & 1 & 1.5 & 1.5 & 0.5 &  0 & 0 & 0 & 0.5 & 0.5 & 0.5 &   8727.158  & -0.015  \\
1 & 0 & 1 & 0.5 & 0.5 & 1.5 &  0 & 0 & 0 & 0.5 & 0.5 & 1.5 &   8727.335  &  0.001  \\
1 & 0 & 1 & 1.5 & 1.5 & 2.5 &  0 & 0 & 0 & 0.5 & 0.5 & 1.5 &   8729.140  & -0.015  \\
1 & 0 & 1 & 1.5 & 1.5 & 1.5 &  0 & 0 & 0 & 0.5 & 0.5 & 0.5 &   8730.424  & -0.013  \\
1 & 0 & 1 & 0.5 & 0.5 & 1.5 &  0 & 0 & 0 & 0.5 & 0.5 & 0.5 &   8741.592  & -0.003  \\
1 & 0 & 1 & 0.5 & 0.5 & 0.5 &  0 & 0 & 0 & 0.5 & 0.5 & 1.5 &   8745.305  &  0.003  \\
2 & 0 & 2 & 2.5 & 2.5 & 2.5 &  1 & 0 & 1 & 1.5 & 1.5 & 2.5 &  17446.707  & -0.017  \\
2 & 0 & 2 & 2.5 & 2.5 & 1.5 &  1 & 0 & 1 & 1.5 & 1.5 & 1.5 &  17455.277  & -0.028  \\
2 & 0 & 2 & 1.5 & 1.5 & 1.5 &  1 & 0 & 1 & 0.5 & 0.5 & 0.5 &  17455.477  & -0.012  \\
2 & 0 & 2 & 2.5 & 2.5 & 1.5 &  1 & 0 & 1 & 1.5 & 1.5 & 0.5 &  17458.542  & -0.026  \\
2 & 0 & 2 & 2.5 & 2.5 & 3.5 &  1 & 0 & 1 & 1.5 & 1.5 & 2.5 &  17458.696  & -0.023  \\
2 & 0 & 2 & 2.5 & 2.5 & 2.5 &  1 & 0 & 1 & 1.5 & 1.5 & 1.5 &  17459.678  & -0.025  \\
2 & 0 & 2 & 1.5 & 1.5 & 2.5 &  1 & 0 & 1 & 0.5 & 0.5 & 1.5 &  17463.217  & -0.008  \\
2 & 0 & 2 & 1.5 & 1.5 & 0.5 &  1 & 0 & 1 & 0.5 & 0.5 & 0.5 &  17463.835  & -0.013  \\
2 & 0 & 2 & 1.5 & 1.5 & 1.5 &  1 & 0 & 1 & 0.5 & 0.5 & 1.5 &  17473.447  & -0.010  \\
2 & 0 & 2 & 1.5 & 1.5 & 2.5 &  1 & 0 & 1 & 1.5 & 1.5 & 1.5 &  17474.380  & -0.004  \\
2 & 0 & 2 & 1.5 & 1.5 & 0.5 &  1 & 0 & 1 & 0.5 & 0.5 & 1.5 &  17481.806  & -0.010  \\
3 & 0 & 3 & 3.5 & 3.5 & 2.5 &  2 & 0 & 2 & 2.5 & 2.5 & 2.5 &  26184.202  & -0.037  \\
3 & 0 & 3 & 3.5 & 3.5 & 2.5 &  2 & 0 & 2 & 2.5 & 2.5 & 1.5 &  26188.605  & -0.032  \\
3 & 0 & 3 & 3.5 & 3.5 & 4.5 &  2 & 0 & 2 & 2.5 & 2.5 & 3.5 &  26188.667  & -0.028  \\
3 & 0 & 3 & 3.5 & 3.5 & 3.5 &  2 & 0 & 2 & 2.5 & 2.5 & 2.5 &  26189.242  & -0.033  \\
3 & 0 & 3 & 2.5 & 2.5 & 2.5 &  2 & 0 & 2 & 1.5 & 1.5 & 1.5 &  26192.521  & -0.022  \\
3 & 0 & 3 & 2.5 & 2.5 & 1.5 &  2 & 0 & 2 & 1.5 & 1.5 & 0.5 &  26192.839  & -0.022  \\
3 & 0 & 3 & 2.5 & 2.5 & 3.5 &  2 & 0 & 2 & 1.5 & 1.5 & 2.5 &  26193.912  & -0.020  \\
3 & 0 & 3 & 2.5 & 2.5 & 3.5 &  2 & 0 & 2 & 2.5 & 2.5 & 3.5 &  26196.613  & -0.004  \\
3 & 0 & 3 & 2.5 & 2.5 & 1.5 &  2 & 0 & 2 & 1.5 & 1.5 & 1.5 &  26201.198  & -0.022  \\
3 & 0 & 3 & 2.5 & 2.5 & 2.5 &  2 & 0 & 2 & 1.5 & 1.5 & 2.5 &  26202.755  & -0.020  \\
3 & 0 & 3 & 2.5 & 2.5 & 3.5 &  2 & 0 & 2 & 2.5 & 2.5 & 2.5 &  26208.613  &  0.000  \\
4 & 0 & 4 & 4.5 & 4.5 & 3.5 &  3 & 0 & 3 & 3.5 & 3.5 & 2.5 &  34918.459  & -0.042  \\
4 & 0 & 4 & 4.5 & 4.5 & 5.5 &  3 & 0 & 3 & 3.5 & 3.5 & 4.5 &  34918.492  & -0.034  \\
4 & 0 & 4 & 4.5 & 4.5 & 4.5 &  3 & 0 & 3 & 3.5 & 3.5 & 3.5 &  34918.844  & -0.037  \\
4 & 0 & 4 & 3.5 & 3.5 & 3.5 &  3 & 0 & 3 & 2.5 & 2.5 & 2.5 &  34923.546  & -0.030  \\
4 & 0 & 4 & 3.5 & 3.5 & 4.5 &  3 & 0 & 3 & 2.5 & 2.5 & 3.5 &  34924.118  & -0.030  \\
4 & 0 & 4 & 3.5 & 3.5 & 4.5 &  3 & 0 & 3 & 3.5 & 3.5 & 4.5 &  34932.067  & -0.003  \\
4 & 0 & 4 & 3.5 & 3.5 & 3.5 &  3 & 0 & 3 & 2.5 & 2.5 & 3.5 &  34932.389  & -0.030  \\
\hline
\end{longtable}
\end{tiny}
\twocolumn

\end{appendix}

\end{document}